\documentclass[12pt]{article}  
\usepackage{amsmath}
\usepackage{url}
\usepackage{graphicx}
\usepackage{verbatim}

\newcommand{\etal}{\textit{et al.\ }}

\usepackage[pagewise]{lineno} 

\usepackage{array}                                            
\usepackage{longtable}                                        
\usepackage{calc}                                             
\usepackage{multirow}                                         
\usepackage{hhline}                                           
\usepackage{ifthen}
\usepackage{lscape}

\begin{document}

\title{Does mutualism hinder biodiversity?\footnote{
Originally submitted on October 2012 as a Brief Communication Arising from James {\em et al.}, {\em Nature} {\bf 487,} 227-230 (2012)}}

\author{Alberto Pascual-Garc\'\i a$^{(1)}$, Antonio Ferrera$^{(2)}$ and
Ugo Bastolla$^{(1,3)}$  \vspace{0.2cm} \\
  \small $^{(1)}$ Centro de Biologia Molecular "Severo Ochoa"\\
  \small CSIC-UAM Cantoblanco, 28049 Madrid, Spain  \vspace{0.1cm} \\
  \small $^{(2)}$ Departamento de Matem\'atica Aplicada y Estad\'\i stica \\
  \small E.T.S.I. Aeron\'auticos, Universidad Polit\'ecnica de Madrid \\
  \small Plaza Cardenal Cisneros 3, Madrid 28040, Spain  \vspace{0.1cm} \\ 
  \small $^{(3)}$ E-mail: ubastolla@cbm.uam.es
}

\date{}

\maketitle
\baselineskip=8.5 mm
\vspace{0.2 in}
 


A recent paper by James \etal$^1$ finds that mutualistic interactions decrease the biodiversity of model ecosystems. However, this result can be reverted if we consider ecological trade-offs and choose parameters suitable for sparse mutualistic networks instead of fully connected networks.
Bastolla \etal$^2$ analytically showed that nested mutualistic interactions reduce the effective competition and increase the structural stability of model ecosystems, making them able to accomodate broader variations of the effective productivity. James \etal$^1$ choose growth rates independently of mutualistic interactions. For sparse mutualistic networks, this procedure increases the variance of the productivity, with negative consequences for biodiversity, consistent with Ref.2.
We propose here how to compare sparse mutualistic networks and fully connected competitive networks in equality of conditions. For both systems, we identify ideal growth rates for which all equilibrium biomasses are equal and the structural stability is maximal, and we perturb these ideal growth rates by the same amount $\Delta$ both for competitive and for mutualistic networks. This procedure incorporates into the model ecological trade-offs, which are consistent with the possible explanation of the stability-complexity debate proposed by May$^3$ and with early field observations$^4$, and necessary for building viable models of empirical mutualistic networks$^2$.

We model obligatory mutualism, with growth rates positive for plants and negative for animals. This makes the model even more challenging and constrains parameters so that the total abundance is $10^4$ to $10^6$ times larger for plants than for animals and the handling times$^5$ are limited. We simulate 500 random realizations of the ecological parameters both for 52 observed mutualistic networks$^2$ and for their purely competitive counterparts (the 4 largest systems in Ref.2 were not studied due to computational limits). The growth rates are centered around the values described in Methods, with relative variance $\Delta$ that represents environmental variability. As expected, the larger $\Delta$ is, the larger the variance of the effective productivity and the more species get extinct. When we compare mutualistic and competitive networks at equal $\Delta$, the relative difference of biodiversity $\delta{S_r}=\left\langle(S_\mathrm{mut}-S_\mathrm{comp})/S_\mathrm{comp}\right\rangle$ is positive, i.e. mutualistic networks support larger biodiversity (Fig.1A).
The handling time parameters $h$ produce a trade-off between the number and the strength of mutualistic interactions, whose importance was noted in Ref.2 (Supplementary Information, pp. 20-23). For broadly distributed mutualistic interactions $h$ cannot be too small, otherwise the trade-off is too weak and mutualism increases the variance of the productivity vector and decreases biodiversity. Nevertheless, the result that mutualism favors biodiversity is robust for a broad range of values of $h$, except very small ones.
Multiple linear regressions using 4 predictors (number of animal and plant species, connectance and nestedness of each network) show that $\delta{S_r}$ is significantly influenced by nestedness (Fig.1B). The influence of connectance is positive and larger than that of nestedness for large $\Delta$, consistent with the results of Ref.1, which correspond to large $\Delta$ and show that connectance influences persistence more than nestedness. Nevertheless, for small $\Delta$ the influence of connectance is negative and that of nestedness is positive.

In summary, the result by James \etal$^1$ that mutualism hinders biodiversity does not hold if we take into account ecological trade-offs and choose parameters in a comparable way both for competitive and for mutualistic networks. If such networks exist in nature, we expect that there should be ecological equations describing their dynamics, possibly similar to those that we propose here.


\newpage

\begin{figure}[hbtp]
\centerline{
\includegraphics[width=0.85\linewidth]{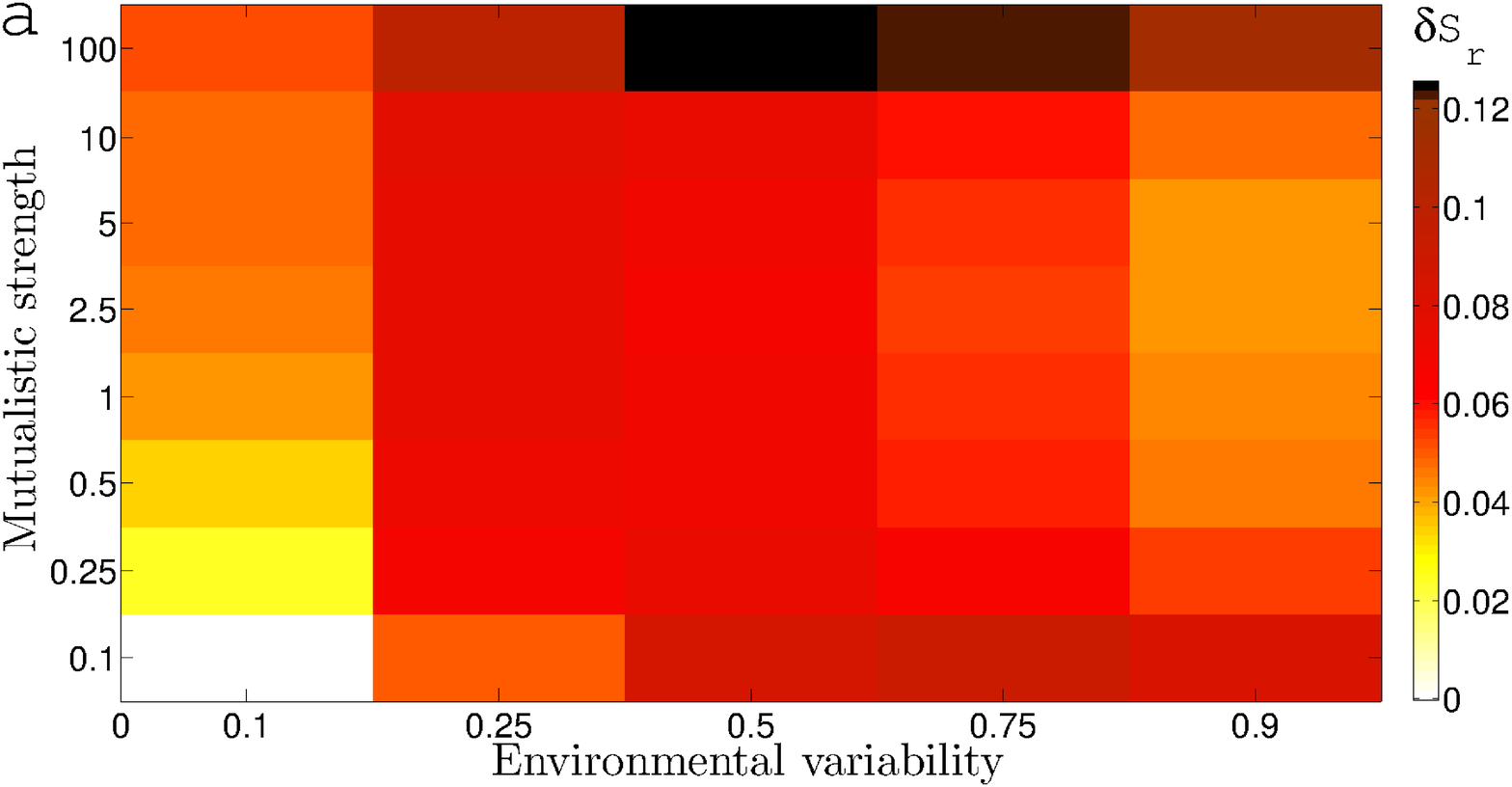}
}
\vspace{0.7cm}
\centerline{
\includegraphics[width=0.72\linewidth]{Comment2James_figureB.eps}
}
\caption{Simulations of obligatory mutualism. Each point represents 52 empirical mutualistic networks, each with 500 realizations of the parameters. (A) Relative difference of biodiversity $\delta{S_r}=\left\langle(S_\mathrm{mut}-S_\mathrm{comp})/S_\mathrm{comp}\right\rangle$ between mutualistic and  competitive networks for different strength of mutualistic interactions $\gamma_0$ and environmental variability $\Delta$. (B) Normalized coefficients of the multilinear regression $Z(\delta{S_r})=A_1{Z(S_A)}+A_2{Z(S_P)}+A_3{Z(\mathrm{connect.})}+A_4{Z(\mathrm{nest.})}$ versus $\Delta$. $Z$ indicates the Z score. Nestedness is quantified through Supplementary Eq.(19) of Ref.2. Data are for $\gamma_0=0.25$. Qualitative results do not change for $\gamma_0\in[0.1,100]$, nor decreasing or increasing the handling times by a factor $1/3$.}
\end{figure}

\newpage

{\large \bf Methods}

We extract the parameters $\alpha^{\mathrm{(P)}}_i$ (growth rates), $\beta^{\mathrm{(P)}}_i$ (competition) and $\gamma^{\mathrm{(P)}}_i$ (mutualism) of Supplementary Eq.(1) of Ref.2 for an empirical mutualistic network with adjacency matrix $\mathcal{A}_{ik}$ as follows:
$\beta^{\mathrm{(P)}}_{ij}=\beta_0\frac{b^{\mathrm{(P)}}_{ij}}{N^{\mathrm{(P)}}}\left[\rho_\mathrm{comp}+\left(1-\rho_\mathrm{comp}\right)\delta_{ij}\right]$, with $\rho_\mathrm{comp}=0.23$, $\gamma^{\mathrm{(P)}}_{ik}=\frac{\gamma_0}{\sqrt{N^{\mathrm{(A)}}N^{\mathrm{(P)}}}}\mathcal{A}_{ik}c^{\mathrm{(P)}}_{ik}$. The dimensionless coefficients $b^{\mathrm{(P)}}_{ij}$ and $c^{\mathrm{(P)}}_{ik}$ are extracted in $[0.85,1.15]$.
$\alpha^{\mathrm{(P)}}_i$ are extracted in $\left[\overline{\alpha}^{\mathrm{(P)}}_i(1-\Delta),\overline{\alpha}^{\mathrm{(P)}}_i(1+\Delta)\right]$, with $\overline{\alpha^{\mathrm{(P)}}}_i=N^{\mathrm{(P)}}\sum_{j\in\mathbf{P}}\beta^{\mathrm{(P)}}_{ij}-\frac{N^{\mathrm{(A)}}\sum_{k\in\mathbf{A}}\gamma^{\mathrm{(P)}}_{ik}}{1+h^{\mathrm{(P)}}_i{N^{\mathrm{(A)}}}\sum_{l\in\mathbf{A}}\gamma^{\mathrm{(P)}}_{il}}$, such that for $\Delta=0$ the abundances $N^{\mathrm{(P)}}_i\equiv\,N^{\mathrm{(P)}}$, $N^{\mathrm{(A)}}_k\equiv\,N^{\mathrm{(A)}}$ are an equilibrium of the dynamical equations.
The handling times are $h^{\mathrm{(P)}}={0.75/\mathrm{max}_i\left(\sum_{j\in\mathbf{P}}\beta^{\mathrm{(P)}}_{ij}\right)N^{\mathrm{(P)}}}$. 
Parameters for animals are obtained interchanging the superscripts P and A.
The time unit is $1/\beta_0=1$, the biomass unit is $N^{\mathrm{(A)}}=1$, $N^{\mathrm{(P)}}$ is the minimum value such that all $\alpha_i$ are positive for plants and negative for animals.
We construct the competitive counterpart as described above, setting $\gamma_0=0$, $\alpha_i>0\;\forall{i}$.

\newpage


\end{document}